\documentstyle[iopfts]{ioplppt}
\begin{document}
\title{Reconstructing the global topology of the universe
from the cosmic microwave background}

\author{Jeffrey R. Weeks}

\address{88 State St, Canton NY 13617, USA}

\begin{abstract}
If the universe is multiply-connected and sufficiently small,
then the last scattering surface wraps around the universe and
intersects itself.  Each circle of intersection appears as two
distinct circles on the microwave sky. The present article shows
how to use the matched circles to explicitly reconstruct the global
topology of space.
\end{abstract}

\section{Introduction}

If the universe is multiply-connected and sufficiently small, then
the last scattering surface (LSS) wraps around the universe and
intersects itself~\cite{css0}.
Each circle of intersection appears as two distinct circles
on the microwave sky, even though the two images correspond
to the same circle of points in space itself.
In their article in this issue, Cornish, Spergel and Starkman~\cite{css1}
show how to find such pairs of matching circles from the high-resolution
data to be provided by NASA's Microwave Anisotropy Probe (MAP)
in the year 2001, or by the ESA's Planck satellite a few years later.
The present article shows how to use the matching circles to
explicitly reconstruct the global topology of space.

The microwave background is isotropic to 1 part in
$10^5$~\cite{bennet}, which implies that the curvature of space is
constant to 1 part in $10^4$~\cite{ratra}.
Our methods work equally well in the spherical, Euclidean, and
hyperbolic cases.  Current evidence suggests space is hyperbolic
with $\Omega_0$ approximately 0.3 or 0.4~\cite{david}.
If $\Omega_0$ is 0.4, the LSS will have a radius of about 2
and enclose a volume of about 75, in units of the curvature radius.
(The curvature radius provides a natural length scale in
hyperbolic as well as spherical geometry.  In spherical geometry
it's usually called a ``radian'', so we will apply that term
in the hyperbolic case as well.).
Thousands of closed hyperbolic 3-manifolds of volume less than 7
are known~\cite{jeff1,jeff2};  each would correspond to a universe
in which the LSS encloses 10 or more images of each object in space,
and in which the topology would be easily detectable.
Moreover, the volume of a hyperbolic 3-manifold is a good measure
of its complexity, and a least action argument~\cite{gary,cgw} suggests
that low-volume universes are more probable than high-volume ones.
We don't rely on such arguments, but they give us hope that
the cosmic topology will be detectable.

\section{Mathematical background}

We will consider space as the quotient of the 3-sphere $S^3$,
Euclidean 3-space $E^3$, or hyperbolic 3-space $H^3$
by a group of covering transformations.
For example, the 3-torus is $E^3$ modulo the group generated by
$x \rightarrow x+1,\; y\rightarrow y+1,\; z\rightarrow z+1$;
a fundamental domain is the cube $0 \leq x,y,z \leq 1$.
By choosing different groups of covering transformations,
exactly ten topologically distinct Euclidean 3-manifolds
may be obtained\cite{hw,bill1}.
Similarly, every spherical manifold may be obtained
as the quotient of $S^3$ by a group of covering transformations.
Infinitely many spherical 3-manifolds are possible, but
they all fall into a few well understood families~\cite{bill1,hopf,st}.
Hyperbolic 3-manifolds, obtainable as quotients of $H^3$,
offer the greatest variety.  Infinitely many are possible,
and their structure is far richer than that of the spherical
manifolds~\cite{bill2}.

\
\begin{figure}[h]
\vspace{70mm}

\includegraphics{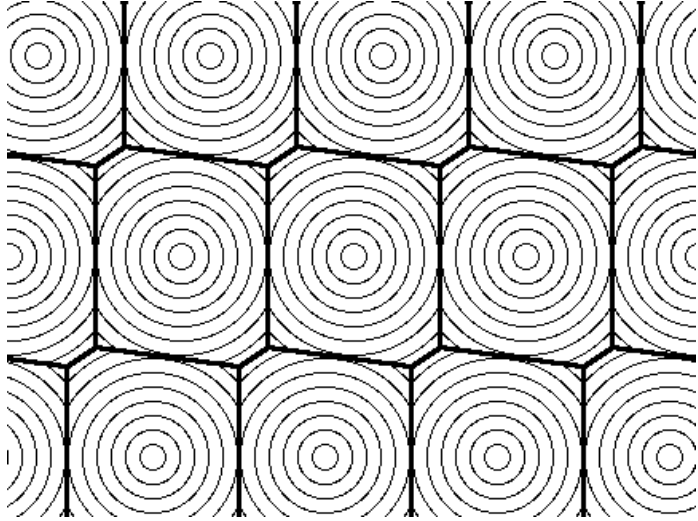}

\vspace{2mm}
\caption{We may ``inflate a balloon'' to construct a fundamental domain
for a closed universe. For clarity the illustration shows multiple
images of the balloon in the universal cover, but the construction is best
imagined in the space itself, where there is only one balloon,
which wraps around the space and presses against itself.}
\end{figure}

The description of a manifold as a quotient of $S^3$, $E^3$, or $H^3$
by a group of covering transformations may easily be converted
to a description as a fundamental domain.
Pick an arbitrary point in the manifold, and start a balloon
expanding at that point (Figure 1).  Eventually the balloon
grows so large that it wraps around the space and touches itself.
When this happens, let the balloon keep expanding.  At the points
where it has touched itself, let it press against itself, forming
a planar disk of contact just as two real balloons (of equal internal
pressure) would form when pressed against one another.
Let the balloon keep expanding until it fills the entire space.
When it has filled the space, the balloon will have the shape of
a polyhedron, with pairs of faces identified to form the original
closed manifold.

\section{Reconstructing the cosmic topology}

We take as our starting point the following data,
all of which may be deduced from the microwave data provided
by the MAP or Planck satellites~\cite{css1}.

\begin{enumerate}

    \item  The geometry of space (spherical, Euclidean or hyperbolic).
    \item  The radius of the last scattering surface.  If the geometry
is spherical or hyperbolic, the radius will be reported in
radians. In the Euclidean case, it will be normalized to 1.
    \item A list of matching circles, as described in the
Introduction.

\end{enumerate}

From these data we will reconstruct the topology of the universe
both as a group of covering transformations of $S^3$, $E^3$, or $H^3$,
and as a fundamental domain.  Let us temporarily assume we are
given perfect data;  after we have laid out the basic algorithm
we will consider possible imperfections in the data, and explain how
to compensate for them.

\
\begin{figure}[h]
\vspace{47mm}

\includegraphics{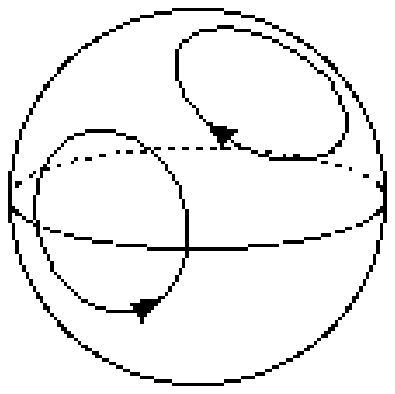}
\includegraphics{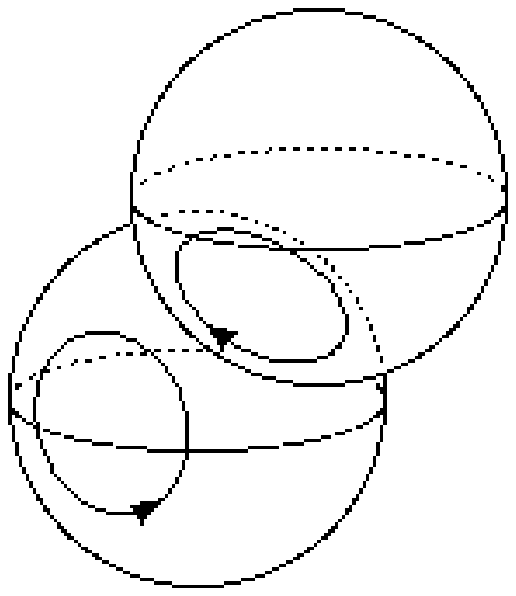}
\includegraphics{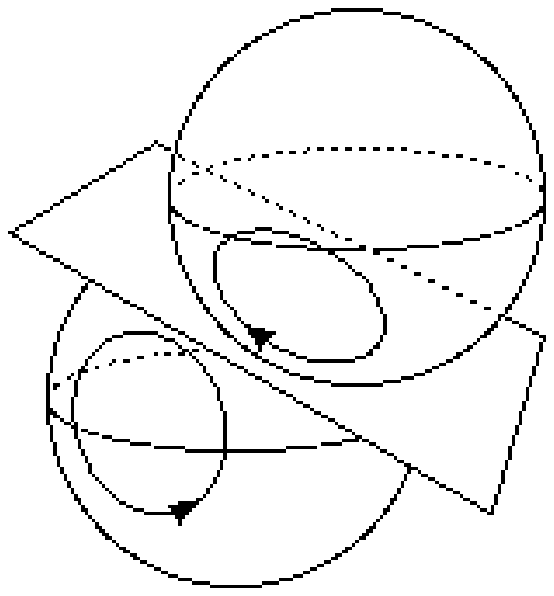}

\vspace{7mm}
\caption{Constructing the fundamental domain}
\end{figure}

\vspace*{-5mm}
\begin{picture}(0,0)
\put(60,40){(a)}
\put(165,40){(b)}
\put(270,40){(c)}
\end{picture}

The main idea is quite simple. Figure 2a shows the LSS, as seen
in the universal covering space. If a pair of circles on the LSS
represent the same circle in the quotient manifold, then there
is a covering transformation $g$ taking one circle to the other.
Figure 2b shows the image of the LSS under the action of $g$.
It is straightforward to compute a matrix for $g$;
in the spherical case the matrix will be in the orthogonal group $O(4)$,
in the hyperbolic case it will be in the Lorentz group $O(3,1)$,
and in the Euclidean case it will be in the subgroup of $GL(4,R)$
fixing the hyperplane $x_0$ = 1.
The transformation $g^{-1}$ interchanges the roles of the two circles.

Constructing a fundamental domain is equally easy. Each matched
circle is equidistant from two images of the observer (Figure 2b).
But each face of the fundamental domain also lies midway
between two images of the observer (Figure 1).
So, roughly speaking, the planes of the circles and the planes
of the fundamental domain's faces coincide! (Figure 2c)
Of course, only the planes of the largest circles correspond to
actual faces of the fundamental domain;  the planes of the smaller
circles are too far away. Conversely, if some face of the
fundamental domain lies entirely outside the LSS, then its plane
does not contain an observable circle;  it lies midway between
two images of the observer, but the corresponding images of the
LSS are too small to intersect. In the extreme case that all
faces of the fundamental domain lie outside the LSS, no circles
are observable and the topology of space cannot be detected.
Assuming for the moment that all faces of the fundamental domain
lie at least partially within the LSS, our algorithm for
constructing the fundamental domain is the following:

\centerline\underline{Algorithm 3.0. Constructing a fundamental
domain.}

\begin{enumerate}
\item Inputs:
\begin{itemize}
\item a space $X$ = $S^3$, $E^3$, or $H^3$,
\item the radius $R_{LSS}$ of the LSS in radians (or 1 in
the Euclidean case)
\item a list $C$ of matching circles
\end{itemize}
\item Output:
\begin{itemize}
\item A polyhedron $D$ (typically a fundamental domain for
the universe -- cf. Proposition 3.1 below)
\end{itemize}
\item Algorithm:
\begin{itemize}
\item  Begin with a ball $B$ of radius $R_{LSS}$ in the
simply-connected space $X$.
\item For each circle $c \in C$, let $P(c)$ be the plane
in $X$ spanned by $c$, and $H(c)$ be the halfspace bounded by $P(c)$
and containing the center of the ball $B$.
\item Let the polyhedron $D$ be the intersection of the halfspaces
$H(c)$, for all $c \in C$.
\end{itemize}
\end{enumerate}

If the microwave data reveal any circles at all, they will probably
reveal a large number of them, and Algorithm 3.0 will compute
a valid fundamental domain for the universe. However, if the
universe is just barely small enough for the LSS to intersect itself,
then we may observe only a few circle pairs, and Algorithm 3.0 may
fail to find all the faces of the fundamental domain.
Proposition 3.1 provides a sufficient condition for checking
that the fundamental domain is correct. Even if the data don't
provide enough circles initially, it's easy to
deduce where the missing faces must lie: compute
matrix generators for the group of covering transformations
(cf. above) and obtain the missing group elements as products
of those generators. The simplest 3-manifolds all have
2- or 3-generator groups~\cite{jeff1}, so 2 or 3 pairs of matched
circles would suffice.

\newtheorem{prop}{Proposition}[section]
\begin{prop}
If the polyhedron $D$ constructed by Algorithm 3.0
lies in the interior of the LSS, then it is a fundamental domain
for the universe.
\end{prop}

\newtheorem{prof}{Proof}[section]
\begin{prof}
Clearly the true fundamental domain must be a subset of $D$.
If the true fundamental domain included a face which $D$ lacked,
then that face would lie in the interior of the LSS.  The plane
spanned by the face would intersect the LSS in a ``matched circle''.
Assuming perfect data (no missing circles), we must have already
found that face. QED
\end{prof}

The circle detection algorithm described in~\cite{css1} is quite reliable.
Nevertheless, we must be prepared for both missing circles
and false matches. Missing circles may be reconstructed by
multiplying together the group elements corresponding to known circles.
False matches may be detected because they won't fit in with the
group structure. That is, the covering transformations corresponding
to all valid circles should fit together to form a discrete group:
the composition of any two such covering transformations should
give another valid transformation. If we find that a few of the
group elements are inconsistent with the overall structure of the
discrete group, then we may reject them and their corresponding
circles as false matches.

The algorithm for constructing fundamental polyhedra has been
implemented as part of the computer program SnapPea, but
only for the hyperbolic case~\cite{jeff1}.
The author is extending it to the spherical and Euclidean cases.
SnapPea lets the user compute a wide variety of invariants for
the resulting manifold, and also check for homeomorphisms
with known manifolds.

\section{Verifying the observational data}

If we do indeed find matching circles in the microwave data,
how can we be certain that our results are correct?
How do we know that the MAP or Planck satellite didn't report
bad data?  How do we know that our computer programs didn't
contain serious bugs?  Fortunately, the discreteness of the group
of covering transformations provides a reliable check against both
observational and computational errors.  The composition of any
two covering transformations must yield a third, to within a
known tolerance.  It is effectively impossible for bad data to
yield a discrete group by chance.  (As mentioned above, some
small portion of the circles may need to be rejected as ``false
matches'', but they are expected to comprise well under 1\% of
the total data. The remaining 99\% of the circle pairs should
then yield a discrete group.)

If we get more than just a few circle pairs,
then we may confirm the data even more dramatically by using
the largest circles to ``predict'' the smaller ones.
That is, we may construct generators for the group
of covering transformations using the largest circles only,
and then take products of the generators to predict the sizes
and locations of all remaining circles.
This method is, of course, equivalent to checking the discreteness
of the group, but most people find it more convincing to
see that a small subset of the data predicts the remainder
of the data set.

\section{Sharpening $R_{LSS}$ and $\Omega_0$}

In the spherical and hyperbolic cases (but not the Euclidean case)
we may use the geometry to sharpen the reported value for the
radius $R_{LSS}$ of the last scattering surface (in spherical or
hyperbolic radians). The sharpened value for $R_{LSS}$ may then
be used to sharpen the values of $\Omega_0$ and other cosmological
parameters.

To sharpen the value of $R_{LSS}$, consider the fundamental domain
computed by Algorithm 3.0. When its faces are identified in pairs
to form the closed manifold, its edges come together in groups
of three. In theory, the sum of the dihedral angles of the three edges
in each group should be exactly $2\pi$. If in practice we find that the
angle sums are all, say, slightly greater than $2\pi$, this implies that
our value of $R_{LSS}$ is too low (in the hyperbolic case)
or too high (in the spherical case). We should replace the old
value of $R_{LSS}$ with a new value which makes the average angle
sum as close to $2\pi$ as possible.

The sharpened value of $R_{LSS}$ may be used to sharpen $\Omega_0$
as well.  The exact relationship between $R_{LSS}$ and $\Omega_0$
depends on the redshift of the LSS, and takes into account
the effects of both matter and radiation.
(For rough estimates, it may be approximated to within a few percent by
$R_{LSS} = {\rm arccos}((2 - \Omega_0)/\Omega_0)$ in the spherical case,
or $R_{LSS} = {\rm arccosh} ((2 - \Omega_0)/\Omega_0)$
in the hyperbolic case\cite{kt}.)
The sharpened value of $\Omega_0$ may in turn be used to sharpen
the values of other cosmological parameters.

\section*{Acknowledgment}

I thank Neil Cornish for patiently teaching me cosmology,
as well as for his generous help in preparing this article.

\end{document}